\documentclass[useAMS,usenatbib,usegraphicx]{mn2e}
\usepackage{epsfig}
\usepackage{amsmath}
\usepackage{amssymb}

\newcommand{\AAA}[3]    {\mbox{#3, A\&A,~#1,~#2}}

\newcommand{\ApJ}[3]    {\mbox{#3, ApJ,~#1,~#2}}

\newcommand{\AJ}[3]     {\mbox{#3, Astron.~J.,~#1,~#2}}
\newcommand{\MNRAS}[3]  {\mbox{#3, MNRAS,~#1,~#2}}
\newcommand{\Nature}[3] {\mbox{#3, Nature,~#1,~#2}}

 \newcommand{\be}{\begin{equation}}
 \newcommand{\ee}{\end{equation}}
 \newcommand{\bea}{\begin{eqnarray}}
 \newcommand{\eea}{\end{eqnarray}}
 \newcommand{\pr}{\partial}
 \newcommand{\Th}{\Theta}
 \newcommand{\N}{\it N \rm}
\begin{document}
\title[Cosmological mass functions and moving barrier models]{Cosmological 
mass functions and moving barrier models}

\author[A. Mahmood and R. Rajesh]
{Asim Mahmood$^{1,2}$
and R. Rajesh$^3$\\
$^1$Astrophysics, Denys Wilkinson building, Keble Road, Oxford, OX1
3RH, UK\\
$^2$ Inter University Centre for Astronomy and Astrophysics, Pune, 411007,
India \\
$^3$Martin Fisher School of Physics, Brandeis University, Mailstop
057, Waltham, MA 02454-9110, USA }
\date{}

\maketitle


\begin{abstract} 
In the ellipsoidal collapse model, the 
critical density for the
collapse of a gravitationally bound object is a function of its mass. 
In the excursion set
formalism, this translates into a moving barrier problem such that the mass
function of dark matter haloes is given by the first crossing
probability of a random walker across
the barrier. In this paper, we study this first crossing probability
analytically. Complete solutions are obtained for barriers that vary as 
square root of time and square of time. Large and small time asymptotic
behaviour is derived. For arbitrary power-law barriers, the large time 
behaviour is
determined. The solutions allow us to derive useful inferences 
about the scaling of the conditional mass function in terms of present day
halo masses and look back redshifts. As an application of our results, 
we compare the
estimates of major merger rates of haloes in constant and moving barrier
models and find that for massive haloes ($10^{12-13} M_\odot$) 
the latter predicts significantly higher merger rates towards high redshifts
($z \gtrsim 4$). 
\end{abstract}

      
\section{Introduction}
\label{sec1}

The problem of determining the mass function of gravitationally bound
structures was first addressed in a successful manner by
\citet{pressschechter}, whose model assumed that the
primordial density fluctuations filtered on a given mass scale were
Gaussian distributed. 
Since their model, many detailed schemes have been investigated and
perfected. The excursion set approach
developed in \citet{bondetal} (hereafter BCEK) and \citet{laceycole1} 
(hereafter LC93) is a convenient tool for deriving the
unconditional and the conditional mass functions within the framework of 
Gaussian random fields. The approach has been successfully 
used to create Monte-Carlo merging history trees of dark matter haloes 
\citep{kauffmannwhiteguiderdoni,somervillekolatt,shethlemson}. 
As compared to \N-body simulations, Monte-Carlo merger trees provide 
alternative faster methods for studying the build-up of dark matter 
haloes. \citet{mowhite} further showed how the
spatial distribution of haloes may be quantified within the 
excursion set approach. 

The excursion set approach is based on the following principles. 
Consider a dark matter inhomogeneity centred
around some point in the universe. The smoothed density contrast
within a radius $R$ around this point is defined as $\bar{\delta}(R) =
[\bar{\rho}(R) - \rho_0]/ \rho_0$, where  $\bar{\rho}(R)$ is the
density of matter within $R$ and $\rho_0$ is the mean
background density of the universe. 
If the density contrast is greater than the
critical density for collapse, the matter contained within the volume
eventually collapses to form a bound object.  
Practically, $\bar{\delta}(R)$ is
obtained by convolving the matter density field with some spherically 
symmetric function $W_R(r)$ of radial extent $R$. 
The variance of the smoothed density contrast is then (e.g., LC93)
\be
S(R) = \langle |\delta(R)|^2 \rangle
= \frac{1}{2\pi^2}\int_0^\infty dk 
\ \langle |\delta_k|^2 \rangle \hat{W}^2_R(k) \ k^2,
\label{variance}
\ee
where $\delta_k$'s are the Fourier amplitudes of the field and 
$\hat{W}_R(k)$ is the Fourier transform of the
window function $W_R(r)$. 

A convenient choice for the window function is the sharp $k$ space 
function, where the cutoff wavenumber $k_s$
is related to the 
Lagrangian radius ($R$) of an object by $R \sim (9 \pi/2)^{1/3}
k_s^{-1}$ (LC93). The equivalent mass scale is $M \sim \rho_0 R^3$. 
The sharp $k$ window function has the advantage that
$\bar{\delta}[R(k_s)]$ executes a random walk
with every increment in the size of the window.
The linearly extrapolated critical density for collapse now serves
as an absorbing barrier for random walk trajectories and the mass
function of collapsed objects
is given by the first crossing distribution of these random
walks across the barrier. 

In the spherical collapse
model, the critical density is independent of
the collapsing halo masses. When linearly extrapolated to the present day,
it has a value of $\delta_c = 1.686$ in a standard Cold Dark
Matter (CDM)
universe. The mass function of dark matter haloes derived using 
the excursion set approach with spherical collapse model
lie within $10-30\ \%$ 
of the results from \N-body simulations \citep{jenkinsetal2}. 
Detailed comparison of simulations with 
analytical results shows discrepancies for both small and 
large mass haloes \citep{gelbbertschinger,laceycole2,
tormen1998,shethtormen1}. 

\citet{shethmotormen} (hereafter SMT01) and \citet{monacoa,monacob}
investigated a non-spherical alternative for collapse of
over-densities. In particular they focused on an ellipsoidal collapse
scenario which derives support from the triaxial nature of
perturbations in Gaussian density fields \citep{doroshkevich,bbks}.
They argued that the main effect of an
ellipsoidal collapse is to introduce a dependence of the critical 
density on the halo mass. In the excursion set formalism this amounts 
to incorporating a `moving' barrier instead of a fixed one. 

In a subsequent paper \citet{shethtormen2} (hereafter ST02) presented a 
detailed discussion on moving barrier models. They considered random walks
in one dimensions $x$ [$\equiv\bar{\delta}(k_s)$] diffusing with time $t$ 
[$\equiv S(k_s)$] starting at $x=0$ when $t=0$. The first crossings with
a moving barrier of the form $B(t)=a+b t^{\gamma}$ were studied.
Based on first crossing distributions $f(t)$ obtained from Monte-Carlo
simulations, they suggested that for a barrier $B(t)$, $f(t)$
has the form 
\be
f(t) = \frac{a e^{-B(t)^2/(2 t)} }
{\sqrt{2 \pi} t^{3/2}} 
\left[ 1 \!+ \!\frac{b t^{\gamma}}{a} \sum_{n=0}^{n^*} \frac{(-1)^n}
{n! (\gamma-n)}
\prod_{k=0}^{n} (\gamma-k)
\right],
\label{raviformula}
\ee
where $n^* \sim 5$. 
In this expression, the diffusion constant $D$ has been
set equal to $1$.
Equation~\ref{raviformula}, when specialised to constant ($b=0$) and
linear barriers ($\gamma=1$), gives the correct answer. For these two
barriers, the first passage distribution is easily obtained by the 
reflection principle \citep{feller} and is
\be
f(t) = \begin{cases}
\frac{a}{t}\frac{1}{\sqrt{2 \pi D t}} \exp\left[\frac{-a^2}{2 Dt}\right]
& b=0, \cr 
\frac{a}{t}\frac{1}{\sqrt{2 \pi D t}} \exp\left[\frac{-(a+b t)^2}{2 D t}\right] 
& \gamma=1.\cr
\end{cases}
\ee
However, the validity of equation \ref{raviformula} for
other kinds of barriers remains unchecked. 
For the two barrier problem, corresponding to the conditional mass function 
(the conditional mass function gives
the progenitor mass distribution for a given present day halo at a given 
look-back redshift), further formulae were suggested based on a
generalisation of the above expression. 

The precise value of $\gamma$ applicable to the ellipsoidal collapse model
seems to lie between $0.5$ and $1.0$. Based on numerics and other
arguments, SMT01 argue that $\gamma\approx 0.6$ with $b>0$. On the
other hand, by applying Lagrangian perturbation theory and considering an 
ellipsoidal collapse model, \citet{monacoa,monacob} concludes
that $\gamma=1/2$ with $b<0$. Also, from the study of tidal torques on galaxy
evolution \citet{delpopoloetal} showed that barrier is of the form $b>0$
and $\gamma \approx 0.58$ \citep{delpopolo}.

In this paper, we present an analysis of the first passage distribution
$f(t)$ for a random walker with a moving barrier. In the process we test the
validity of equation \ref{raviformula}. We present
an analytical solution of the square root barrier ($\gamma=1/2$). This
barrier is close to the one studied in SMT01 ($\gamma \approx
0.6$). We also solve for the
quadratic boundary ($\gamma=2$). The large and small time behaviour of these
solutions are also derived. For arbitrary $\gamma$, we calculate the large
time behaviour of the first crossing probability. Our results show that
equation \ref{raviformula} is not correct in general and needs to be modified.

We also present the 
methodology for approaching the problem of conditional 
mass functions in the context of moving barrier models. 
The conditional mass function requires solving
a two barrier problem. 
Section~\ref{sec2} describes how to modify the moving barrier problem so that
both the conditional as well as unconditional mass distributions may be
obtained. 

The rest of the paper is organised as follows.
In section~\ref{sec3}, we present the analytical solution for the square root
barrier ($\gamma=1/2$). The large and small time asymptotic behaviour is
derived. Expressions are also obtained for conditional mass distributions.
The analytical results are compared with results from Monte-Carlo simulations.

In section~\ref{sec:quadratic}, we consider another solvable limit -- the
quadratic barrier ($\gamma=2$). The full solution and the large time
asymptotic behaviour are derived, and compared with results from simulations.
For both these cases, we show that the results are not consistent with the
formula in equation~\ref{raviformula}.

In section~\ref{sec:arbitrary}, we consider barriers with arbitrary $\gamma$.
For this general case, we argue what the large time behaviour of $f(t)$ should
be. Using special algorithms, we numerically obtain $f(t)$ for large times
and confirm our prediction.

In section \ref{sec:nbody} we fit the halo mass
distribution obtained from $N$-body simulations with the square root barrier
results. While a good fit is obtained, the numerical data is not good enough
to differentiate between different values of $\gamma$.

In section~\ref{sec4}, we describe an application of the results to estimate
major merger rates of haloes. These rates are calculated using the 
conditional mass functions for the square root barrier.
These rates are compared with those derived from a constant barrier
Press-Schechter model. The analysis suggests that while the two models
are very similar in describing the low redshift evolution of these
rates, there is a systematic deviation 
to wards high redshifts. In particular the cumulative
major merger rates of massive haloes ($10^{12-13} M_\odot$) differ
significantly at redshifts $z \gtrsim 4$.

We conclude with a summary and discussion in
Section~\ref{sec6}. In Appendix~\ref{appendixa}, results for parabolic
cylinder functions that are relevant to the paper are reproduced.

Unless otherwise stated, the cosmological
parameters used will be $\Omega_m = 0.3$, $\Omega_\Lambda = 0.7$, $h =
0.7$ and root mean square fluctuations at $8\ h^{-1} \rm Mpc$ as
$\sigma_8 = 0.9$. We also interchangeably use $x \leftrightarrow \delta$ and
$t \leftrightarrow S$.

\section{The moving barrier model}
\label{sec2}

For the ellipsoidal collapse model, the critical density
contrast depends on the variance or `scale' $S$ and has the form 
\citep{shethmotormen}
\be
\delta_{ec}(S,z) = \sqrt{q}\ \delta_{sc}(z)\left[ 1 + \beta 
\left( \frac{S}{q  S_*} \right)^\gamma \right]. 
\label{barrier1}
\ee
Here, $\delta_{sc}(z)$ is 
the spherical collapse critical density at redshift $z$ and
$S_* = \delta_{sc}^2(z)$. The best-fit values of parameters $\beta$, $\gamma$
and $q$ for the mass function of haloes in GIF simulations of the Virgo 
consortium \citep{jenkinsetal1} are $0.5$, $0.6$ and $0.707$
respectively.  The value of $q$ 
depends on the way haloes are identified in the simulations and
therefore is a function of the link-length in the halo finding
algorithms.

The variable $\delta$ executes a random walk when $S$ is
increased. We are interested in the probability that $\delta$ exceeds the
critical density $\delta_{ec}$ at scale $S$.
In the following analysis it is convenient to substitute
the variables $\delta$ and $S$ by variables $x$ and $t$ respectively.
Consider a random walk starting at $x=0$ when $t=0$. Let $P(x,t)$ be the
probability of finding it at $x$ at time $t$. Then, $P(x,t)$ obeys the
diffusion equation
\be
\frac{\pr P(x,t)}{\pr t} = \frac{D}{2}\frac{\pr^2 P(x,t)}{\pr x^2},
\label{diffusioneq}
\ee
where $D$ is the diffusion constant with a value one in the present
context. In terms of $x$ and $t$, the barrier in equation \ref{barrier1}
takes the form $a+b t^\gamma$, where
$a,b$ are constants.

In what follows, we will consider modified barriers of the form
\be
B(t) = a+b (t+t_0)^\gamma,
\label{barrier}
\ee
where $t_0$ is some constant (see below 
for motivation for this choice). Let $g(a,b,t_0;t)$ be the probability that 
the random walk crosses
the barrier $B(t)$ for the first time at time $t$. The unconditional mass
distribution is obtained by setting $t_0=0$. To calculate the conditional mass
function 
one has to consider a two barrier problem.

\begin{figure}
\includegraphics[width=\columnwidth]{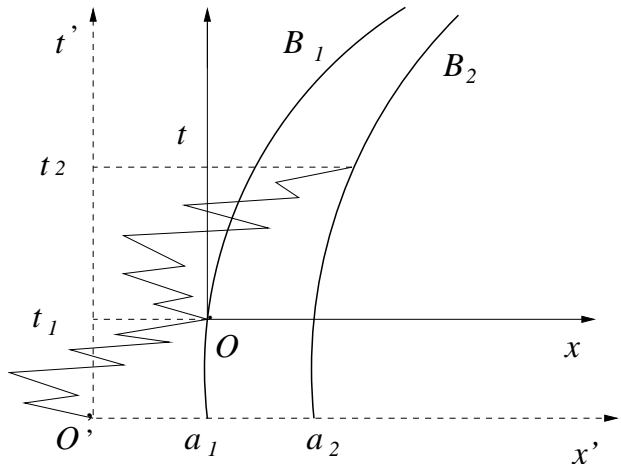}       
\caption{Illustration of the two barrier problem corresponding to the
conditional mass distribution.}
\label{randomwalk}
\end{figure}

Consider the illustration in figure \ref{randomwalk}. The
random walk trajectories begin at position $O'$. The moving barrier 
$B_1$ represents the critical density at present epoch $z_1$
and has the form $B_1 \equiv a_1 + b_1  t'^\gamma$.  
At an earlier epoch $z_2$ the
critical density is represented by barrier $B_2 \equiv a_2 + b_2
t'^\gamma$. For earlier look-back epochs or redshifts (increasing $z$), 
$a_2>a_1$.  The precise dependence of
$a$ on $z$ is determined from the cosmological model. For example,
in an $\Omega_m =1$ universe $a_{1,2} = 1.686\times  (1 + z_{1,2})$ [Note:
the parameter $a$ should not be confused with the scale factor which is
often represented by the same notation].

The jagged curve shows a random walk trajectory that
meets the barrier $B_1$ for the first time at some scale $t_1$ and the
barrier $B_2$ for the first time at some `smaller' scale $t_2$. 
This represents a collapsed object of scale $t_1$ at 
redshift $z_1$ that had a collapsed (or `formed') progenitor of scale 
$t_2$ at redshift $z_2$.
Let $f(t_1\ \cap \ t_2)$ be the probability of this occurring.
To determine $f(t_1\ \cap \ t_2)$, we need to know the probability 
that a random walk
starting at the point $O$ reaches the barrier $B_2$ for the first time in
time $t_2-t_1$. In the new coordinates $x,t$, the barrier takes the form $a+b
(t+t_1)^\gamma$, where $a=a_2-a_1-b_1 t_1^\gamma$ and $b=b_2$. Thus,
\be
f(t_1 \ \cap \ t_2)=g(a_1,b_1,0;t_1)\ g(a_2-a_1-b_1 t_1^\gamma,b_2, t_1;
t_2-t_1).
\label{eq:conditional}
\ee
The conditional mass distribution $f(t_2|t_1)$ is then given by
\be
f(t_2\ |\ t_1) = g(\Delta a-b_1 t_1^\gamma,b_2, t_1;t_2-t_1),
\label{eq:conditional1}
\ee
where $\Delta a = a_2-a_1$. 
Thus, a knowledge of $g(a,b,t_0;t)$ solves both the conditional as well as
unconditional mass distributions. 
This was the motivation for introducing the
parameter $t_0$ into the modified barrier.
In what follows, we will be considering the
case when $b_1=b_2=b$.

The first step in solving
equation \ref{diffusioneq} is to make the boundary condition 
(equation \ref{barrier}) time independent by
choosing suitable coordinates. Let
$x \rightarrow B_1(t) - x$, with $t$ remaining unchanged.
Then, equation \ref{diffusioneq} reduces to
\be
 \frac{\partial P(x,t)}{\partial t} 
+ b  \gamma  (t + t_0) ^{\gamma -1} \frac{\partial P(x,t)}{\partial x}
- \frac{D}{2} \frac{\partial^2 P(x,t)}{\partial x^2} = 0,
\label{diffeq}
\ee
satisfying the boundary conditions
\bea
P(0,t)  & = & 0, \label{boundarycond1} \\
P(\infty, t) & = & 0, \label{boundarycond2} \\
P(x, 0) & = & \delta_D  [  x - (a + b  t_0^\gamma) ] \label{boundarycond3},
\eea
where $\delta_D$ is the Dirac delta function. The first
crossing distribution is 
\bea
g(a,b,t_0;t) &=& - \frac{\pr}{\pr t} \int_0^{\infty} P (x,t) \ d
x.  \label{firstcrossing1}\\
&=& \frac{D}{2} \left[ \frac{\pr P}{\pr x} \right]_{x=0},
\label{firstcrossing}
\eea
where the integral in equation \ref{firstcrossing1} has been evaluated 
using equation \ref{diffeq}.

We now bring  equation~\ref{diffeq} to a more convenient form by the
following transformations.
Let $P(x,t) = \phi(x,t) e^{h(x,t)}$.  Substituting into equation
\ref{diffeq}, we determine the function $h(x,t)$ by eliminating
the terms proportional to $\pr \phi / \pr x$. 
Thus,  if we choose $h(x,t)$ to be 
\be  
h(x,t) = \frac{b \gamma x (t+t_0)^{\gamma -1}}{D} - \frac{b^2 
\gamma^2 (t+t_0)^{2 \gamma -1}}{ 2 D (2 \gamma -1)}, \ \gamma \neq
\frac{1}{2}, 
\ee
then equation \ref{diffeq} simplifies to
\be
\frac{\pr \phi}{ \pr t} +  \frac{b \gamma (\gamma-1) x}{D
(t+t_0)^{2-\gamma}} \phi
- \frac{D}{2} \frac{\pr^2 \phi }{\pr x^2}=0, \ \gamma\neq \frac{1}{2}.
\label{phidiffeq}
\ee

When $\gamma=2$, the second term in equation~{\ref{phidiffeq} becomes
independent of time, and the equation becomes separable. We present the full
solution for this case in section~\ref{sec:quadratic}. Also, it turns out
that when $\gamma=1/2$, the equation~\ref{diffeq} can be made separable 
by a different set of transformations. We present the complete solution for
this case in section~\ref{sec3}. For arbitrary $\gamma$, it is not possible
to transform equation~{\ref{phidiffeq} into a separable form. However, it is
possible to derive some results in the limit of large time. We discuss this
in section~\ref{sec:arbitrary}.

\section{The square root barrier}
\label{sec3}

In this section we solve for the first crossing probability for the square
root barrier ($\gamma=1/2$).

\subsection{\label{sec3.1} Solution}

Let $y = x/\sqrt{t + t_0}$ and $\eta = t$.
Then equation \ref{diffeq} simplifies to 
\be
(\eta+t_0) \frac{\pr P}{\pr \eta} +
\frac{b - y}{2}\frac{\pr P}{\pr y}  -
\frac{D}{2}\frac{\pr^2 P}{\pr y^2} = 0.
\label{eq:separate}
\ee
Equation \ref{eq:separate} is now separable.
Let $P(y,\eta) = \Theta(\eta)  \Psi(y)$. Then,
\bea
\frac{(\eta +t_0)}{\Theta(\eta)} \frac{\partial \Th}{\partial \eta} &
= & - \lambda  \\
\frac{D}{2  \Psi}\frac{\partial^2 \Psi}{\partial y^2} + \frac{(y -
  b)}{2 \Psi} 
\frac{\partial \Psi}{\partial y}  & = & + \lambda,
\label{eigeneq}
\eea
where $\lambda$ is an eigenvalue. The first of these equations is
easily integrated to give
\be
\Th(\eta) =\frac{1}{(\eta+t_0)^{\lambda }}.
\ee 

In equation \ref{eigeneq}, we substitute $(y-b)/\sqrt{D} = \zeta$. Then,
\be
\Psi(\zeta)'' +  \zeta \Psi(\zeta)' + 2 \lambda\Psi(\zeta) = 0.
\label{psieq}
\ee
Further, letting $2 \lambda -1 = v$, and
applying a transformation 
$\Psi(\zeta) = \Phi (\zeta) \exp (-\zeta^2/4)$, 
we obtain
\be
\Phi''(\zeta) + (v + \frac{1}{2} - \frac{1}{4}\zeta^2 ) \Phi(\zeta) = 0. 
\label{pcdiffeq} 
\ee
The solutions to this equation are the parabolic cylinder functions
$U_v(\zeta)$ and the eigen values $v$ are determined by the boundary
condition $U_{v} (-b/\sqrt{D}) = 0$ (from equation \ref{boundarycond1}). 
For large $\zeta$, the function $U_v(\zeta)$ goes
to zero as $\zeta^v \exp(- \zeta^2/4)[1 + O 
(1/\zeta)]$, consistent with the other boundary condition. 

It is interesting to note a similarity between the present problem and
the problem of a quantum particle trapped in a potential $V(\zeta)$ given
as  
\be
V(\zeta)
= \begin{cases}
\infty, & \zeta \leq -\frac{b}{\sqrt{D}}, \cr 
\frac{1}{4} \zeta^2, & \zeta > -\frac{b}{\sqrt{D}}.\cr
\end{cases}
\ee
In the limit $b=0$, the problem reduces to the harmonic oscillator potential
with an $V(\zeta) =\infty$ for $\zeta \leq 0$. The parabolic cylinder functions then
reduce to Hermite polynomials with
$U_{2n + 1}(z) = 2^{-n-1/2 }H_{2n + 1}(z)  e^{-z^2/4}$.

For finite $b$ exact eigen-values can be obtained numerically. 
Alternatively, the large $n$ behaviour of $v_n$ may be obtained from the
large $v$ asymptotic behaviour of the 
parabolic cylinder function (see appendix~\ref{appendixa}). Then,
\be
v_n = 2n  - \frac{2 \sqrt{2}  b}{\pi \sqrt{D} }\sqrt{n} +
\frac{2  b^2}{\pi^2D} +(2 k-1) +
O\left(\frac{1}{\sqrt{n}}\right),~n\rightarrow \infty,
\label{eigenval}
\ee
where $k$ is an integer.

The probability $P(x,t)$ can now be written as a
linear combination of all eigen modes as
\bea
P (x,t) &=& \sum_n \frac{A_{v_n}}{(t+t_0)^{(v_n +1)/2}}
U_{v_n}\left(\frac{x-b \sqrt{t+t_0}}{\sqrt{D (t+t_0)}}\right) \nonumber \\
&& \mbox{} \times \exp \left[- 
\frac{(x-b \sqrt{t+t_0} )^2 }{ 4 D (t+t_0)}\right].
\eea
Here $A_{v_n}$ are constants that are determined through the initial
condition. Using the orthogonality relation for parabolic cylinder 
functions 
\be
\int_{-b/\sqrt{D}}^{\infty} U_{\mu}(\zeta) U_v(\zeta) d \zeta  =  0, 
\ \mu \neq v \label{orthog}.	
\ee
and using the initial condition \ref{boundarycond3}, one obtains
\be
A_{\mu} = \frac{t_0^{ \mu/2} e^{a^2/(4 D t_0)}}
{\sqrt{D} \ I_{\mu}(-b/\sqrt{D})}  
U_{\mu}\left(\frac{a}{\sqrt{D t_0}} \right),
\label{amu}
\ee
where we have defined $I_\mu (-b/\sqrt{D})$ as
\be 
I_{\mu}\left(\frac{-b}{\sqrt{D}} \right) = \int_{-b/\sqrt{D}}^{\infty} 
U_{\mu}^2(\zeta) \ d \zeta.
\label{imu}
\ee

Having obtained $P(x,t)$, we can derive the distribution $g(a,b,t_0;t)$ 
in a straightforward manner using equation \ref{firstcrossing}:
\bea
\lefteqn{
g(a,b,t_0;t)=
\frac{e^{-b^2/(4D)} e^{a^2/(4 D t_0)}}{2(t+t_0)} \times}  \nonumber \\
&& \sum_{\{ v \}} \left( \frac{t_0}{t+t_0} \right)^{v/2} 
\frac{U'_v(-b/\sqrt{D})U_v(a/\sqrt{D t_0})}{I_{v}(-b/\sqrt{D})},
\label{cfcd}
\eea
where $U'_v(x)$ denotes a first derivative
with respect to $x$. 
The conditional first crossing distribution can be
obtained by using equation \ref{eq:conditional1}.
To obtain the unconditional distribution 
we take the limit $t_0 \rightarrow 0$ in 
equation \ref{cfcd} to obtain
\be
g(a,b,0;t) = \frac{\exp(-b^2 / 4D)}{2 t} \sum_{v} \left( \frac{a^2}{D t}
\right)^{v/2} \frac{U'_v(-b/\sqrt{D})}{I_v(-b/\sqrt{D})}. 
\label{fcd}
\ee
It is straightforward to show that one recovers the constant barrier answer
from this expression by setting $b=0$.

In figure \ref{fig_comp}, we compare the analytical result 
for unconditional mass
function (equation \ref{fcd}) with results from Monte-Carlo simulations.
In terms of the barrier form given in equation \ref{barrier1}, the
parameter values are $\gamma = 0.5$, $\beta = 0.5$ and $q =
1$. In the
four panels (i, ii, iii and iv), $a=\delta_{sc}(z) = 1,2,3$ and $4$ respectively
The solid curve is obtained by truncating
the series in equation \ref{fcd} to the first $30$ terms. 
There is a marked
difference between the moving barrier and the constant barrier especially for
large times (or small masses).
\begin{figure}
\includegraphics[width=\columnwidth]{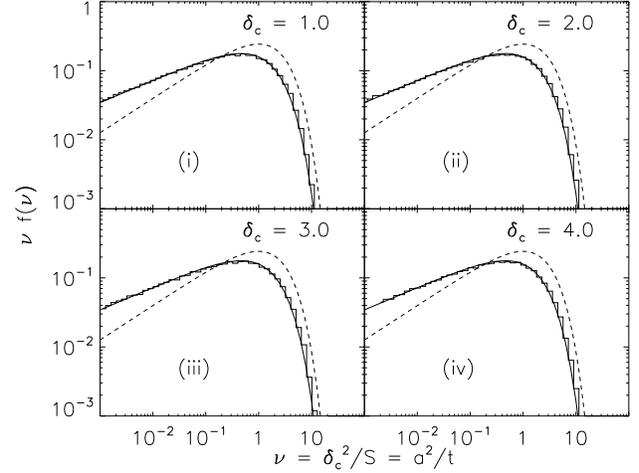}       
\caption{The unconditional mass function or the first
crossing distribution for $\gamma=1/2$ (solid line) is compared with results
from Monte Carlo simulations (histogram) for different values of $\delta_c$.
The dashed curve is the mass function obtained from the 
Press-Schechter constant barrier model.}
\label{fig_comp}
\end{figure}

We also compare the expression for the square root barrier with Monte Carlo
simulations for a $\gamma=0.6$ barrier. This is shown in 
figure~\ref{fig_comp2} where the solid curve is the expression in equation
\ref{fcd} and the histogram is data for $\gamma=0.6$. The other parameters are
same for both simulation and the curve. It is seen that the data lies close
to the curve. In fact, if the parameters $b$ and $a$ were allowed to be
played with, a much better fit can be obtained. This shows the difficulty in
distinguishing between various $\gamma$'s by looking at the mass
distribution.
\begin{figure}
\includegraphics[width=\columnwidth]{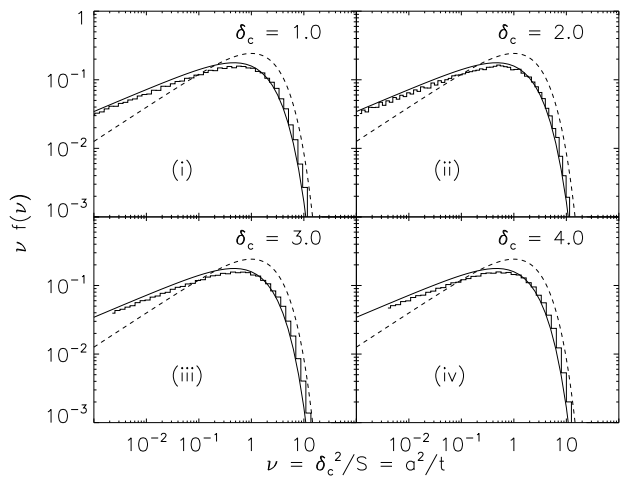}       
\caption{
Same as in figure~\ref{fig_comp} except that the histogram is now for
$\gamma=0.6$.  }
\label{fig_comp2}
\end{figure}

\subsection{\label{sec3.2}Large time behaviour of the unconditional 
first crossing probability}

In this subsection, we consider the large time behaviour of the
first crossing probability. In the large time limit, the leading order
contribution arises from the first term in the infinite series in equation
\ref{fcd}. 
This corresponds to the variable $\nu = a^2 /( D t) \rightarrow 0$
limit.
The unconditional first crossing distribution behaves as $f(t)\sim
t^{-\theta}$ with $\theta= 1 + v_0/2$, where $v_0$ is the smallest
eigenvalue. 
It is not possible to
analytically calculate $v_0$ for arbitrary $b$. 
When $b\rightarrow \infty$, the eigenvalue problem reduces to the
harmonic oscillator problem with lowest eigenvalue being $0$. Thus one
expects $\theta\rightarrow 1$ as $b\rightarrow \infty$. When $b\rightarrow
-\infty$, the barrier crosses the walker almost immediately. Thus, one
expects that $\theta\rightarrow \infty$ when $b\rightarrow -\infty$.
For small $b$ ($b^2 \ll D$), using the
variational method, \citet{krapivskyredner} obtained an approximate value
of the exponent as $\theta \approx 3/2 - b/\sqrt{2 \pi D}$. 
At the same time, one could naively assume that the large $v$
behaviour for parabolic cylinder functions (see equation 
\ref{parabolic_asymp}) is valid for all $v$ and then 
estimate $v_0$. Under this assumption,
one obtains
\be
\theta = \frac{3}{2} - \frac{b}{\pi^2 D} \sqrt{b^2 + \pi^2 D} +
\frac{b^2}{\pi^2 D}.
\label{theta_eq}
\ee
Surprisingly, this expression agrees very well with the real answer and also
has the right limits for large absolute values of $b$.
Figure \ref{theta} plots the behaviour of the
exponent $\theta$ as a function of $b/\sqrt{D}$.
This scaling is markedly different from the fixed barrier 
scaling in large time limit $f(t) \propto t^{-3/2}$. 
\begin{figure}
\begin{center}
\includegraphics[width=50mm]{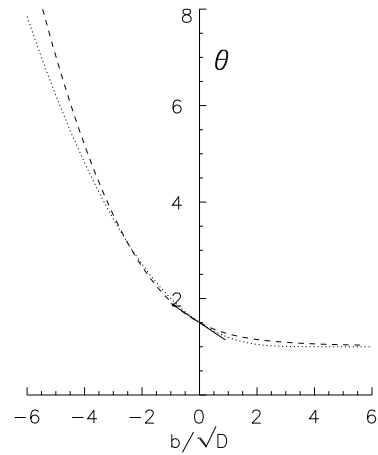}       
\caption{The variation of the exponent $\theta$ with $b/\sqrt{D}$ is shown
for the square root barrier.
The dotted line is obtained by numerically
solving $U_v(-b/\sqrt{D}) = 0$ for $v$; the solid line shows the
exponent obtained by \citet{krapivskyredner} and the dashed line is
from equation \ref{theta_eq}. As expected, when $b/\sqrt{D} =0$, the
value of the exponent is $3/2$. }
\label{theta}
\end{center}
\end{figure}

We mention here that, if we were to look at the large time behaviour of
$f(t)$ in equation \ref{raviformula} with $\gamma=1/2$, then 
$\theta =3/2-\gamma=1$. Clearly, the lack
of dependence on the parameter $b$ makes it qualitatively different from the
actual answer.

\subsection{\label{sec3.3}Small time behaviour of the unconditional 
first crossing probability}

In the small time limit [$\nu=a^2/(Dt) \gg 1$], we can determine the
asymptotic behaviour of the first crossing probability $g(a,b,0;t)$ by the
saddle point method. In this limit, the leading contribution to
$g(a,b,0;t)$ comes from a term with large $v$ in the
infinite series of equation \ref{fcd}. 
On substituting the large $v$ asymptotic form of
functions $U'_v$ and $I_v$ (see Appendix A), and replacing
the summation by an integral over $n$ by noting from equation
\ref{dn} that $dn = [ 1/2 + b/(2 \pi\sqrt{v D} )  ]  d v$, equation
\ref{fcd} reduces to
\bea
f(t) &=& \frac{\exp(-b^2/4 D)}{2 \pi t \sqrt{2}} \int_0^{\infty}  
d v \left\{ \left(\frac{a^2}{Dt}\right)^{v/2} 
\left(\frac{e}{v}\right)^{v/2} \right.  \nonumber \\
&&  \times \left. \sin\left(\frac{\pi v}{2} + 
\frac{b \sqrt{v}}{\sqrt{D}} \right)   [ 1 + 
  O(1/v) ]  \right\}.  
\label{fta}
\eea
Here we have only retained terms up to $O(1/\sqrt{v})$.
Rewriting the `sine' term as an exponential, and
evaluating the integral
by the saddle point method, we obtain,
after some algebra, the small time asymptotic form of the
expression as
\bea
f(t) &= &\frac{a}{\sqrt{2 \pi D} t^{3/2}} \exp\left[-\frac{(a + b
\sqrt{t} )^2}{2 D  t}\right] \nonumber \\
&&\mbox{} \times \left[ 1 + \frac{b(b^2 + 6D) \sqrt{t}}
{24 D  a} +O\left(\frac{D t}{a^2}\right) \right].
\label{fexp}
\eea

Comparing this expression with equation \ref{raviformula}, it appears that
the latter should be considered as an expansion for the small time behaviour
of $f(t)$. Specialising to $\gamma=1/2$, $f(t)$ has the same form as in
equation \ref{fexp} except for the coefficient in front of the correction
term. But whether this qualitative similarity continues to exist for
$\gamma>1/2$ is not very clear. In fact, in sections~\ref{sec:quadratic} and
\ref{sec:arbitrary}, we will show
that equation \ref{raviformula} does not have the right form for arbitrary
$\gamma$.

\subsection{\label{sec3.4}Asymptotic analysis of the conditional mass function}

In this subsection, we study the asymptotic behaviour of the conditional
mass distribution. This is given by equation~\ref{eq:conditional1}.
Using the result for $g(a,b,t_0;t)$ (see equation~\ref{cfcd}), we obtain
\bea
\lefteqn{
f(t_2\ |\ t_1)= \frac{1}{2  t_2} \exp\left[
\frac{(\Delta a)^2} {4 D t_1} 
- \frac{b \Delta a} {2 D \sqrt{t_1}} 
\right]
\times}  \nonumber \\
&& \sum_{\{ v \}} \left( \frac{t_1}{t_2} \right)^{v/2} 
\frac{U'_v(-b/\sqrt{D})U_v[(\Delta a - b \sqrt{t_1})/\sqrt{D t_1}]}
{I_{v}(-b/\sqrt{D})},
\label{eq:final_cond}
\eea
where $\Delta a = a_2-a_1$. Unlike the fixed barrier case, the conditional
mass function is no longer a universal function of just one scaling
variable $\nu=\Delta a/\sqrt{t_2 - t_1}$. 

In figure~\ref{fig_cond_n}, 
the conditional mass function as obtained from equation
\ref{eq:final_cond} (solid curve), is plotted with the results from Monte-Carlo
simulations (histograms) for a range of present day halo masses ($M_1$) and
look-back redshifts. Here $M_*$ is such that $S(M_*) = \delta_{sc}^2(0)$,
which, for $\Lambda$CDM cosmology used here, has a value $M_* \sim 2
\times 10^{13} M_{\odot}$.
The plot confirms the correctness of our expression. The
result from the constant barrier model is shown for comparison (dashed
curve).
In figure \ref{fig_cond} we show the effect of
rescaling the conditional mass function in terms of variable $\nu$. As
expected, the shape of the curve is different at different look-back
epochs and for different present day halo masses.
\begin{figure}
\includegraphics[width=\columnwidth]{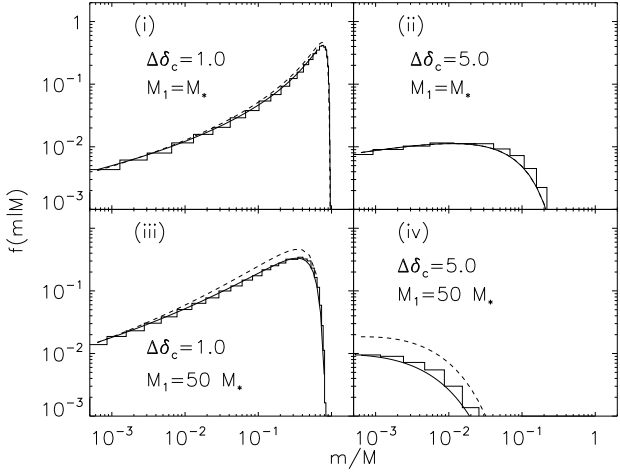}       
\caption{The analytical conditional mass function (solid curve) is
compared with the results from
Monte-Carlos simulations (histograms). The dashed curve is the fixed
barrier conditional mass function. In panel (ii), the two
curves are not distinguishable from each other. 
Here $M_1$ is such that $t_1 = S(M_1)$.}
\label{fig_cond_n}
\end{figure}
\begin{figure}
\includegraphics[width=\columnwidth]{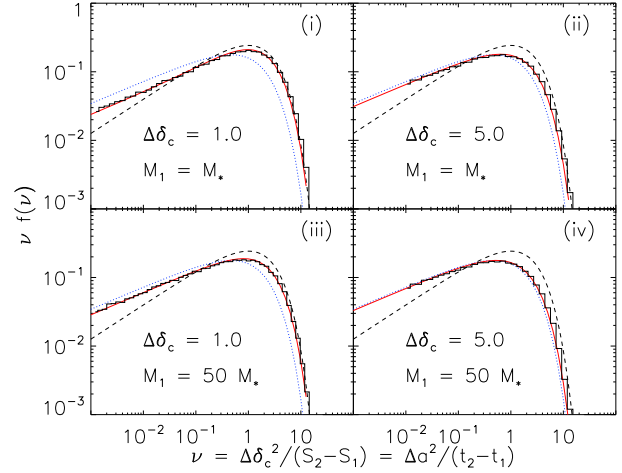}       
\caption{The conditional mass function plotted in terms of variable
  $\nu = \Delta a^2 /(t_2 - t_1)$. The histograms show the results of 
Monte-Carlos simulations. The dotted curve shows the distribution
obtained by re-scaling of the unconditional mass function.
The dashed curve is the constant barrier mass function.}
\label{fig_cond}
\end{figure}

We will consider the following limiting cases. Consider the limit $(\Delta a
/\sqrt{D t_1}) \rightarrow \infty$. This is the same limit as the one that was
taken to obtain the unconditional mass distribution. Doing so, one obtains
\be
\lim_{\frac{\Delta a}{\sqrt{D t_1}} \rightarrow \infty} f(t_2|t_1) = 
\frac{e^{-b^2 / (4 D)}}{2 t_2} \sum_{v} \left( \frac{\Delta a^2}{D t_2}
\right)^{v/2} \frac{U'_v(-b/\sqrt{D})}{I_v(-b/\sqrt{D})},
\ee
which is the same as equation~\ref{fcd} with $a$ replaced by $\Delta a $.
Therefore, for a given look back epoch, the
conditional mass function tends to the unconditional mass function for
large present epoch masses. Alternatively, for a fixed present epoch
mass, the conditional distribution tends to the unconditional
distribution at large look back epochs. 

We now consider the opposite limit
$(\Delta a/\sqrt{D t_1}) \rightarrow 0$. Taking the Taylor series
expansion of the function  $U(\Delta a/\sqrt{Dt_1}- b/\sqrt{D})$ we can write
\be
U_v\left[ \frac{\Delta a - b \sqrt{t_1} }{\sqrt{D t_1}} \right]
\approx  U_v(-b/\sqrt{D}) + \frac{\Delta a}{\sqrt{D t_1}}
U'_v(-b/\sqrt{D}), 
\ee
where have only taken terms up to order $\Delta a /\sqrt{D t_1}$. 
Note that the first term on the right hand side is zero from the 
boundary condition. Thus equation \ref{eq:final_cond} can be written as
\be
f(t_2\ |\ t_1) = \frac{\Delta a}{\sqrt{D t_1}} 
\frac{1}{2  t_2 } \sum_{\{ v \}} \left( \frac{t_1}{t_2} \right)^{v/2} 
\frac{U'_v(-b/\sqrt{D} )^2}{ I_{v}(-b/\sqrt{D})}. 
\label{asympcond}
\ee
In this limit there does not appear to be
a direct way of discerning the behaviour of
$f(t_2|t_1)$. 
However, figure \ref{fig_cond} suggests that 
the form of the conditional mass function in this limit 
is closer to the constant barrier mass function. This point has also been discussed
in ST02: for a given $t_1$, at small look back epochs 
most random walkers will reach the second
barrier in a relatively short time so that the barrier hasn't moved
much and is effectively fixed. ST02 argue that the effect should be
more pronounced for massive haloes. Instead,
panels (i) and (iii) of figure \ref{fig_cond} suggest
that for small mass haloes 
the conditional distribution are more like the
constant barrier distribution, whereas for more massive haloes, the
conditional distribution tends to the unconditional distribution.
Within the context of the present analysis, this may be understood by
noting that for a given $\Delta a$, as $M_1 \rightarrow \infty$
or equivalently $t_1 \rightarrow 0$, equation \ref{eq:final_cond} approaches
the unconditional mass function. A simple argument can also be put forward
by looking at the barrier form under consideration: $B \equiv \Delta a
- b (t_1^\gamma - t_2^\gamma )$ [see section \ref{sec2}]. 
When $t_2 \rightarrow t_1$ (note that the minimum value of $t_2$ is $t_1$) 
the situation is similar to the fixed barrier case. On the other hand
when $t_2 \gg t_1$, 
the situation is similar to the single 
barrier problem. Therefore, for a given present halo mass and look
back redshift, one expects a transition from a fixed barrier
distribution to the unconditional moving barrier distribution.

\section{The quadratic barrier}
\label{sec:quadratic}

In this section, we solve for the first passage distribution for the
quadratic barrier ($\gamma=2$). We start with equation~\ref{phidiffeq}, which
when specialised to $\gamma=2$ is
\be
\frac{\pr \phi}{ \pr t} +  \frac{2 b x}{D} \phi
- \frac{D}{2} \frac{\pr^2 \phi }{\pr x^2}=0.
\label{phidiffeq2}
\ee
Let $\phi(x,t) = \psi(x) T(t)$. Then,
\bea
\frac{1}{T}\frac{\pr T}{\pr t} = - \lambda, \\
\frac{2  b}{D} x - \frac{D}{2  \psi} \frac{\pr^2 \psi}{\pr x^2} = +
\lambda, \label{eq:quadratic}
\eea
where $\lambda$ is an eigenvalue.

The time dependent part has a solution $T(t) = e^{-\lambda
t}$. Let $y = k(2  b 
x/D  - \lambda) $, where 
\be
k = [D/(2 b^2) ]^{1/3}.
\ee
Then equation \ref{eq:quadratic} reduces to the Airy equation
\be
\frac{\pr^2 \psi}{\pr y^2} - y \psi = 0.
\ee
The solution to the above equation are Airy functions $Ai(y)$ such
that when $y \rightarrow \infty$, $Ai(y) \rightarrow 0$. The other
boundary condition $\psi(x=0)=0$ implies 
$Ai(-k \lambda) = 0$ and this fixes 
the eigen values $\lambda_n$. Hence, we can write solution in the form
\be
\phi(x,t) = \sum_{n = 1}^{\infty} C_n e^{-\lambda_n t }Ai\left[ k \left(
\frac{2  b  x}{D} - \lambda_n\right)\right],
\label{eq:phieq}
\ee
where the coefficients $C_n$ are  to be determined from the initial
condition.

The function $P(x,t)$ can now be written as
\be
P(x,t)\! = \! \sum_{n =1}^{\infty}\! C_n \exp\!\! \left[ - \lambda_n t\! +\! 
\frac{2 b x t}{D} \!  - \! \frac{2 b^2 t^3}{3 D} \right] 
\!\!Ai \!\left[k \left\{\!\frac{2  b  x}{D}\! -\! \lambda_n \!\right \}\right].
\ee
The initial condition now gives us
\be
\delta_D(x - a) = \sum_{n=1}^{\infty} C_n Ai \left[  k \left(\frac{2 
b x}{D} - \lambda_n\right)\right].
\label{parabolicinit}
\ee
Multiplying both sides of this equation by $Ai \left[  k \left(\frac{2 
b  x}{D} - \lambda_m\right)\right]$ and integrating between the
boundaries $x= 0$ and $x = \infty$, we obtain, using the orthogonality
of Airy functions,
\be
Ai \left[  k \left(\frac{2  b  a}{D} - \lambda_m\right)\right] =
C_m \int_0^\infty dx  Ai^2 \left[  k \left(\frac{2  b x}{D} -
\lambda_m\right)\right].
\ee
The integral on the right hand side can be evaluated using the
relation $Ai''(x) = x Ai(x)$ and is equal to: $\frac{D}{2  b  k}
Ai'^2(-k \lambda_m)$, where $Ai'$ is the derivative of the Airy
function. Thus we obtain the coefficients as
\be
C_m = \frac{2 b k}{D} \frac{1}{Ai'^2(-k \lambda_m)} Ai\left[  k
  \left(\frac{2  b a}{D} -  \lambda_m\right)\right].
\ee

Now using equation \ref{firstcrossing} we can write the expression for
the first crossing distribution as
\be
f(t) = \frac{2 b^2  k^2 e^{-2 b^2 t^3/(3 D)}}{D} 
\sum_{n=1}^{n=\infty} e^{- \lambda_n t}
\frac{Ai(2 b a /D-\lambda_n)}{Ai'(-k \lambda_n) }.
\label{gamma2ft}
\ee
Furthermore, the asymptotic form of the function $Ai(-x)$ for $x
\rightarrow \infty$ is given as
\be
Ai(-x) = \frac{1}{\sqrt{\pi} x^{1/4}} \sin \left(\frac{2}{3} x^{3/2} +
\frac{\pi}{4}\right), ~x\rightarrow \infty.
\ee
Thus using $Ai(-k \lambda_n) = 0$, we obtain the eigen values as
\be
\lambda_n = \frac{1}{k} \left[ \frac{3 \pi}{2} \left( n - \frac{1}{4} \right) 
\right]^{2/3}, \ n\rightarrow \infty.
\ee

In the large time limit $t \rightarrow \infty$ it can be seen that the
leading order term in the first crossing distribution is $f(t)
\propto \exp[- 2 b^2 t^3 /(3D)]$. This can be compared with the 
expression in equation \ref{raviformula}:
$f(t) \propto \exp[- b^2 t^3 /(2D)]$. Thus, when $\gamma
=2$, the expression by ST02 is wrong.
Figure \ref{fig_gamma2}
shows the first crossing distribution for $\gamma = 2$ for fiducial parameter
values: $b = 0.2$ and $a = 1.675$. The
solid curve shows
the result from equation \ref{gamma2ft} 
and the histograms show the
distribution obtained from Monte-Carlo simulations.
\begin{figure}
\includegraphics[width=\columnwidth]{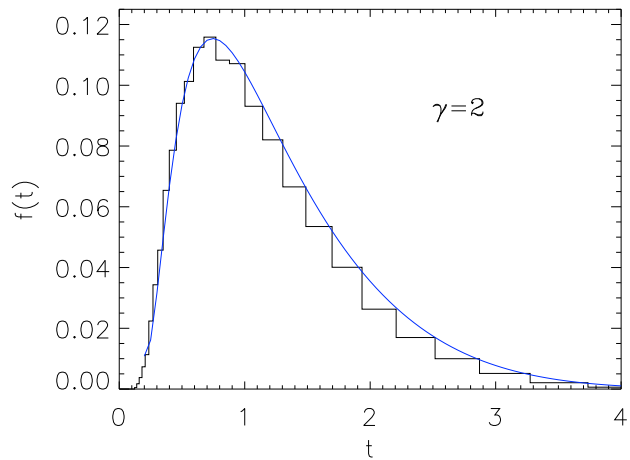}       
\caption{The unconditional first crossing probability for a 
quadratic barrier. The
fiducial values of parameters are $b = 0.2$ and $a =
1.675$. 
The solid curve is from equation \ref{gamma2ft} and
histograms are from Monte-Carlo simulation of random walks.}
\label{fig_gamma2}
\end{figure}

\section{Arbitrary \Large{$\gamma$}}
\label{sec:arbitrary}

For $\gamma$ different from $0,1/2,1,2$, it is not possible to obtain the
full solution for $f(t)$. However, it is possible to analyse it in the limit
of large $t$. This analysis would be useful to test whether equation
\ref{raviformula} has the right form or not. In this section we estimate
\be
\alpha = \lim_{t\rightarrow \infty} \frac{\ln (f(t)}{t^{2 \gamma-1}},
\ee
such that $f(t) \propto \exp(\alpha t^{2 \gamma-1})$. For comparison, equation
\ref{raviformula} gives $\alpha=-b^2/2$.

We start with equation \ref{phidiffeq} with $t_0=0$. $\phi(x,t)$ obeys the
boundary condition $\phi(0,t)=0$. The first passage distribution $f(t)$ is
then given by
\be
f(t) = \frac{D}{2} \left. \frac{\partial \phi}{\partial x}\right|_{x=0} 
\exp\left[\frac{-b^2 \gamma^2 t^{2 \gamma-1}}{2 D (2 \gamma-1)}\right].
\label{eq:49}
\ee
We will argue that the contribution from $ \partial \phi/\partial
x|_{x=0}$ is at an order much smaller than the term in the exponential in
the limit $t\rightarrow \infty$.

For large $t$, one could treat the potential in equation \ref{phidiffeq} as a
slowly varying linear potential. We then make the adiabatic approximation. As
$t\rightarrow \infty$, we expect the system to be in ground state of the time
dependent linear potential. Then, $\phi(x,t)$ would be
approximately equal to the first term in equation \ref{eq:phieq} with $b$
replaced by $b \gamma (\gamma-1) t^{\gamma-2}/2$. Thus, one would expect that
the contribution from $\phi$ to $f(t)$ is utmost of the order $\exp[t^{(2
\gamma -1)/3}]$. From equation \ref{eq:49}, we then conclude that
\be
\alpha= \frac{-b^2 \gamma^2 }{2 D (2 \gamma-1)}.
\label{eq:alpharesult}
\ee

We now compare the above heuristic result for $\alpha$ with 
results from numerical simulations (with $D=1$). It turns out that it
is more convenient to numerically measure the survival probability
$S(t)$ rather than directly measure $f(t)$. $S(t)$ is the probability that a
random walker has not crossed the barrier up to time $t$ and is related to
$f(t)$ by $f(t)=-dS(t)/dt$; hence it has the same $\alpha$ as $f(t)$.
To estimate $\alpha$ we need to go to large times, but then face the problem
that $S(t)$ becomes exponentially small. It is difficult to overcome
this problem using conventional Monte Carlo methods. Instead,
we use an algorithm known as ``go with the winners'' 
algorithm \citep{grassberger}. We briefly describe the algorithm.
We start with $N$ ($N=2\times 10^5$ in our case) random walkers
at the origin. When the number of walkers get reduced by half (due to
absorption at the boundary), copies are
made of the remaining surviving ones, and the survival probability is halved. 
By repeating this procedure, we can go down to very low values of survival
probability keeping the number of live walkers constant. As an example
of the kind of data that is obtained, we show the survival
probability for $\gamma=0.75$ in figure \ref{fig_survival}.
\begin{figure}
\includegraphics[width=\columnwidth]{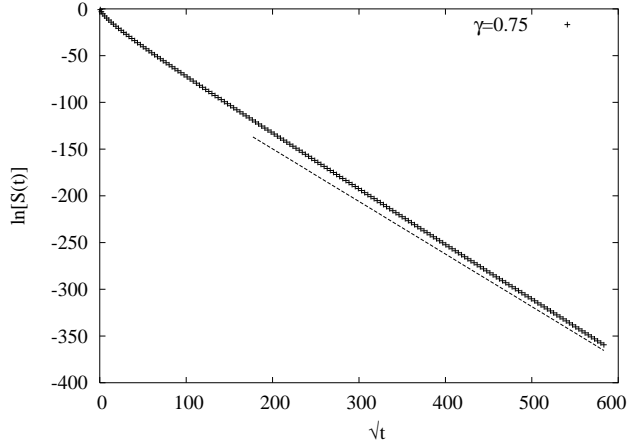}       
\caption{The variation of the logarithm of the 
survival probability $S(t)$ with $\sqrt{t}$
is shown. The dotted line has slope as in
equation~\ref{eq:alpharesult} and is shown for comparison. 
Very small $S(t)$ can be reached 
using specialised algorithms.}
\label{fig_survival}
\end{figure}

From the numerical results for the survival probability, 
$\alpha$ is determined by best
fit.  In figure \ref{fig_alpha} we show the variation of $\alpha$ with $\gamma$.
The data agrees well with the formula in equation
\ref{eq:alpharesult}. We would expect that equation
\ref{eq:alpharesult} is an exact result.
\begin{figure}
\includegraphics[width=\columnwidth]{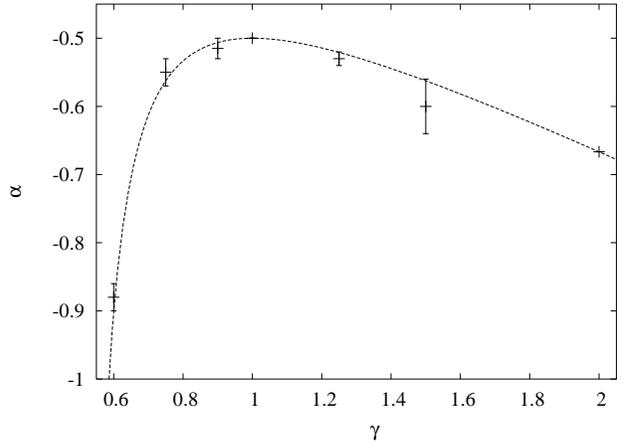}       
\caption{The variation of $\alpha$ with $\gamma$ is shown. The values for
$\gamma=1$ and $\gamma=2$ are exact. The data is for $b=1.0$. The solid line
is the formula is equation \ref{eq:alpharesult}.}
\label{fig_alpha}
\end{figure}

\section{Halo Mass function from {\it N}-body simulations}
\label{sec:nbody}

A two parameter empirical formula for the halo mass function as
given in \citet{shethtormen1} can be written as
\be
\nu f(\nu) = A \sqrt{\frac{q \nu}{2 \pi}} \left[1 +  (q \nu)^{-p}\right] \exp
\left( - \frac{q \nu}{2}\right).
\label{nbody_mf}
\ee
Here $\nu = \delta_{sc}(z)^2/\sigma^2(m)$. 
The parameters are $q =0.707$, $p =0.3$ and $A\approx 0.322$, which is set by
enforcing the condition that integral of $f(\nu)$ over all $\nu$
equals unity. The Press Schechter formula has $q =1$, $p =0$ and
$A=1/2$. For small $\nu$ one obtains $\nu f(\nu) \propto
\nu^{0.5-p}$. Equivalently for large $t$, $f(t) \propto t^{3/2-p}$.

In this section we fit the {\it N}-body mass function with the unconditional
mass function for the square root barrier, taking $b$ and $q$ as free
parameters (see equation \ref{barrier1}). For this, we use the
small time asymptotic form of the unconditional mass function as given
in equation \ref{fexp}. We use the halo catalogues of GIF simulations
for $\rm \Lambda CDM$ cosmology, as available on
$http://www.mpa-garching.mpg.de/Virgo/$. The
cosmological parameters for the simulation are $\Omega_m = 0.3$,
$\Omega_\Lambda = 0.7$, $h =0.7$ and $\sigma_8 = 0.9$. The box size of
these simulations was $L=141\ h^{-1} \rm Mpc  $ and a total of $256^3$
particles were simulated. We only consider haloes with at least 50
particles and a maximum mass of $3 \times 10^{14}M_\odot$. This upper
limit is chosen as the statistics of haloes with masses $\gtrsim 3 \times
10^{14} M_{\odot}$ is dominated by significant noise. 
For lower halo mass limit one should
typically consider $\gtrsim 70$ particles (Arif Babul private
communication).
The best-fit value of the parameters we obtain are $b \approx 0.5$ and $q
\approx 0.55$. This value of $q$ is somewhat smaller than that
obtained by fitting the expression \ref{nbody_mf}. 
Interestingly, using the large time behaviour of
$f(t)$ as discussed in section \ref{sec3.2}, for $b=0.5$ one obtains
$f(t) \propto t^{3/2-p'}$, where $p'\approx 0.2$. In the following
analysis we restrict to our best-fit values for $q$ and $b$.

In figure \ref{massfn_root}, the left panel shows the GIF mass
function (filled circles) and the solid curve shows the mass function
as given in equation \ref{fcd}. Also plotted for
comparison is the Press Schechter mass function (dashed curve) and the
asymptotic form of mass function as given by equation \ref{fexp}
(dotted curve). In the figure we
have labelled the square root barrier mass function as `ELL' as in
ellipsoidal collapse. The fact that the full solution gives a worse fit than
the asymptotic formula (with the same parameters) shows the danger in 
fixing parameters through approximate formulae.

The
right panel shows the \citet{shethtormen1} mass function as given by
equation \ref{nbody_mf}. It can be seen that the square root barrier
mass function gives a reasonable description of the halo mass function
obtained from the simulations. It also shows the difficulty of extracting the
correct $\gamma$ from the simulations. To do so, one needs to have good data
for small mass, and the resolution at these mass scales is not good enough.
\begin{figure}
\includegraphics[width=\columnwidth]{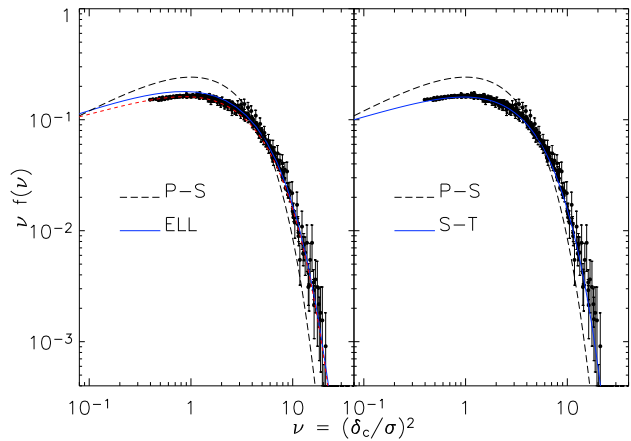}       
\caption{The halo mass function. Filled circles show the mass function
obtained from GIF simulations. The error bars show the poissonian
error. \emph{Left panel:}  The dashed curve shows the Press Schechter
mass function and the solid curve shows the moving barrier mass
function (see section \ref{sec:nbody} for details). The
dotted curve shows the asymptotic form of the unconditional mass
function (equation \ref{fexp}). \emph{Right panel:} The solid curve
shows the best-fit empirical mass function obtained in
\citet{shethtormen1}.}
\label{massfn_root}
\end{figure}

\section{Implications for merging history of haloes}
\label{sec4}

In this section, we present an application of our calculation. We use the
results of the conditional mass distribution for the square root barrier to
estimate the major merger rates of dark matter haloes. In simple
models \citep{wyitheloeb02,wyitheloeb,mahmoodetal},
halo mergers are followed by galaxy mergers. Major
mergers of galaxies are thought to trigger off quasars.
Thus halo merger rates
provide an estimate of quasar numbers at any given redshift. 
This is particularly useful, since quasars can be observed directly up
to sufficiently high redshifts ($z \gtrsim 6$).
Previous studies
have shown that the Press-Schechter (PS) halo major merger rates are
consistent with the evolution of quasars at high redshifts ($z \gtrsim
2$) but fail for low redshifts \citep{wyitheloeb02}.
The low redshift evolution of quasars is marked by 
an enhanced role of baryonic components and is therefore probably 
beyond the reach of simple predictions based on halo major merger rates. 

In the PS model the simplest estimates of halo `creation' rates are 
often taken as the positive term in the cosmological-time (or
redshift) derivative of the
unconditional halo mass function (we explicitly write
cosmological-time so as to avoid any confusion with the notation
$t,t_0,t_1,t_2$ used in the paper).
This is partly a necessity as in
the PS formalism all haloes are continuously accreting matter and are
therefore newly formed. There is no direct way of
differentiating between minor and major accretion episodes. To counter
this deficiency of the model, a simple argument is used to define a
major merger: the haloes which accrete mass comparable to their own
mass, in a short interval of time are said to undergo a major merger. 
Using the conditional mass function, therefore, one can compute the
probability that at a very small look back time $\Delta a$, 
a given halo had a `formed' progenitor less than half of its present
mass. Substituting $\Delta a$ by 
$\rm d a/ d \tau$ (where $\tau$ is the
cosmological time), in this probability yields the rate at
which haloes of a given present day masses are forming through major
mergers. We call these rates as major merger rates of haloes.
This approach has been discussed, for e.g., in \citet{mahmoodetal}. 

Alternatively instead of looking back-wards one can compute the 
probability that a given halo merges with a comparable mass 
object (`formed' or otherwise) during a small step $\Delta a$ 
forward in time \citep{wyitheloeb}. The two approaches yield
essentially similar results.

\subsection{\label{sec4.1}`Creation' rates}

We first consider the implications of taking the positive term in the 
cosmic-time derivative of the unconditional mass function. 
The left panel in figure \ref{cum_rate} shows the redshift evolution of
creation rates as obtained from the Press Schechter mass function
(dashed curve) and 
the mass function for the square root barrier (solid curve). These rates
have been integrated over the minimum masses as depicted in the figure.
It can be seen that the moving barrier mass function gives
considerably higher creation rates for large redshifts.
\begin{figure}
\includegraphics[width=\columnwidth]{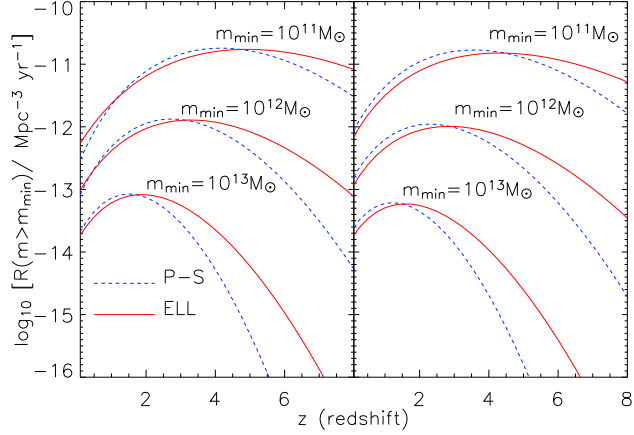}       
\caption{\emph{Left Panel:} The cumulative creation rates of haloes with
minimum masses as depicted in the figure. \emph{Right panel:} The
cumulative major merger rates of haloes, above some given masses
(see section \ref{sec4.2} for details).}
\label{cum_rate}
\end{figure}

\subsection{\label{sec4.2}Major merger rates}

In order to compute major merger rates of haloes, we consider an
object with a given $t_1$ (corresponding to mass $M_1$) at some epoch; for
a small look back time $\Delta a \rightarrow 0$, we can apply the
result of equation \ref{asympcond} to obtain $f(t_2 | t_1)$.
The fraction of objects with less than half mass progenitors 
can therefore be computed as
\bea
\lefteqn{ w(t_2 > t_h|\ t_1) = \int_{t_h}^\infty d t_2 f(t_2|t_1)} \nonumber \\
&&=  \frac{\Delta a}{\sqrt{D  t_1}}\sum_{v}
\frac{1}{v} \left( \frac{t_1}{t_h} \right)^{v/2}
\frac{U'_v(-b/\sqrt{D})^2}{ I_{v}(-b/\sqrt{D})},
\eea
where $t_1 \equiv S(M_1)$ and $t_h \equiv S(M_1/2)$. Replacing $\Delta
a$ by $\dot{a} = \rm da/d\tau$ 
gives us $\dot{w}$. This represents the fraction
of objects of mass $M_1$ formed from major mergers, per unit time. 
Multiplying $\dot{w}$ with the existing number of objects of mass
$M_1$ as determined from the mass function, gives the `major merger
rate' of haloes of mass $M_1$. Integrating the rate over $M_1$ from some
minimum mass $M_{min}$ to $\infty$ gives the cumulative major merger
rate of objects above mass $M_{min}$.

In the right panel of figure 
\ref{cum_rate} we plot cumulative major merger rates of haloes in the constant barrier 
(dashed curve) Press Schechter model and the moving barrier (solid curve) model. 
For the moving barrier model 
the unconditional mass function is as given by equation \ref{fexp}.
Figure \ref{cum_rate} shows that the moving barrier model yields a higher
number of cumulative major merger rates towards high redshifts and particularly
so for massive haloes ($\sim 10^{12-13} M_{\odot}$). This is precisely
the halo mass range which is relevant for quasars. The
inclusion of moving barrier threshold could,
therefore, significantly affect the predictions of analytic and
semi-analytic models dealing with the evolutionary history of the
high redshift quasars. 

As an example we compute the number counts of
quasars above a given redshift and flux level in observed soft 
$(0.5-2 \rm \ keV)$ and hard $(2-10 \rm \ keV)$ X-ray bands.
For this, we use the
model discussed in \citet{mahmoodetal}. The black hole mass ($M_{BH}$)
to halo mass ($M_h$) relation is as described in that paper. The major
merger rates of haloes are computed as discussed above. Using the
$M_{BH}-M_h$ relation these rates are converted into black hole
formation rates. The bolometric luminosities are estimated as
the Eddington luminosity for a given black hole mass. To obtain the
luminosity in observed X-ray bands, at a given redshift $z$,
we derive bolometric corrections
in bands $0.5(1+z)-2(1+z) \rm \ keV$ and $2(1+z)-10(1+z) \rm \
keV$ respectively. 
For this we use the spectral energy distribution and the rest-frame
bolometric corrections described in \citet{marconietal04}. 
The quasar life times are taken as the local dynamical times of
galaxies (note that the relative difference between the Press
Schechter and the moving barrier model will be independent of the
lifetime). Thus luminosity functions in $0.5(1+z)-2(1+z) \rm \ keV$
and $2(1+z)-10(1+z) \rm \ keV$ bands are obtained.

Quasar luminosities are related to the observed flux as $F_X(L_X,z) =
L_X/4\pi/D_L(z)^2$ (here $L_x$ is the X-ray luminosity). From the
luminosity function $\phi(L_X,z)$ of quasars, the number of
X-ray quasars in the whole sky, above some flux level $F_X$ and above
some redshift $z$ is given
as \citep{haimanloeb}
\be
N(> F_X,>z) = 4 \pi \int_{z}^{\infty} \!\!\! dz' 
\! \int_{L(F_X,z)}^{\infty}
dL_X \left( \frac{d^2V}{dz' d \Omega}\right) \phi(L_X,z).
\ee
Here $d^2 V/dz/d\Omega$ is the comoving volume element per unit
redshift and per unit solid angle. The left panel of figure
\ref{flux_plot} shows the number count in a $6'$ circle, for hard 
X-ray band. The
right panel depicts the results for soft X-ray band. The solid curve
is the moving barrier model prediction and the dashed curve is the Press
Schechter prediction. It can be seen that there is a pronounced
discrepancy between the two models at high redshifts. Even though the
full nature of this discrepancy can only be known
through a model based on {\it N}-body merger trees, our results
highlight the difference qualitatively. It may be added here that the
difference in the models primarily owes to the difference in the halo
mass function in the two models.
\begin{figure}
\includegraphics[width=\columnwidth]{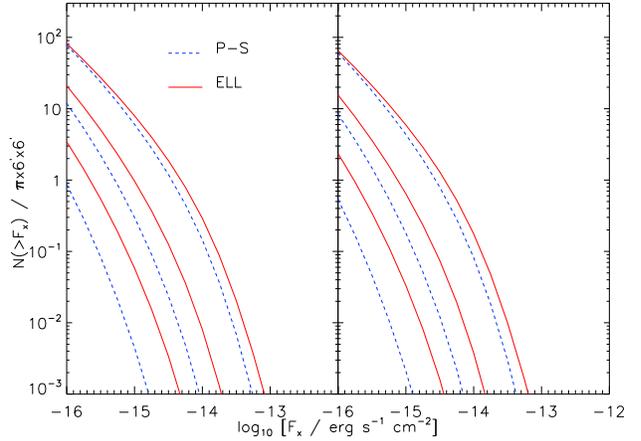}       
\caption{\emph{Left panel:} The number count of quasars in the
  observed hard X-ray band
($2-10 \ \rm keV$). In each line style, the set of three curves from
  top to bottom correspond to $z \gtrsim 3,5$ and $7$ respectively. \
  Solid curves show the moving barrier prediction
  and dashed curves show the prediction of Press Schechter
  model. \emph{Right panel:} The number count of quasars in the
  observed soft X-ray
  band ($0.5-2 \ \rm keV$).}
\label{flux_plot}
\end{figure}

\section{Summary and discussion}
\label{sec6}

In this paper we studied the first
crossing probability $f(t)$ of one dimensional random walks across a barrier
moving as $t^{\gamma}$. 
Complete analytical solutions for
the square root barrier ($\gamma=1/2$) and the quadratic barrier ($\gamma=2$) were presented. For arbitrary power law barriers, 
large time estimates of the first crossing probability were derived.
We showed that the formula
for $f(t)$ as presented in \citet{shethtormen2} is not
valid for general $\gamma$. We also presented a
methodology for approaching the two barrier problem for conditional
mass function of haloes. This is relevant
for deriving the progenitor distribution of haloes. Unlike the
stationary and the linear barrier, the conditional first crossing
probability does not follow a simple re-scaling of the unconditional
first crossing probability for other barrier forms. 

In \citet{shethmotormen}, it was argued that $\gamma \approx 0.6$ is the
barrier arising from ellipsoidal collapse. It was also shown the mass
function from $N$-body simulations could be fitted well by a barrier of this
kind. In this paper, we fitted the $N$-body data using the square root
barrier. This barrier has the advantage of being analytically tractable and
the conditional distributions are also fully known. As shown in
section~\ref{sec:nbody}, the numerical data is fitted well with the square
root barrier too. Since we also derived large time behaviour of first passage
distributions across barriers with
arbitrary $\gamma$, one could ask the following question. Treating $\gamma$
as a free parameter, can it be determined using the data from $N$-body
simulations? For small times (large masses), $f(t)$ is dominated by the
stationary barrier. It is only for large times that $\gamma$ plays a role.
Thus, to determine $\gamma$, the $N$-body simulations should have good
resolution at small masses. This is currently not available, and it is not
possible to determine $\gamma$ from the simulations with any degree of
confidence.

We also compared the predictions for merger rates of
haloes in the constant barrier model with that of the moving barrier model. 
The moving barrier model predicts a significantly higher
merger rates for massive haloes towards high redshifts. This 
arises as a consequence of the fact that the moving barrier
mass function yields more massive
haloes at high redshifts, as compared to the standard Press Schechter
mass function. 
In this regard an interesting exercise would be to
compare the halo merger rates obtained from simulations with the
constant and the moving barrier merger rates.
In the present
context it appears that the prediction of the moving barrier model is
consistent with the early formation of massive galaxies
\citep{glazebrooketal} and the presence of high redshift quasars
with massive host haloes \citep{fanetal}. We also  presented a
calculation of X-ray quasar number count above a given flux
level and a given redshift. In terms of relative abundances, we found
that the moving barrier model predicts systematically more quasars
towards higher redshift. At $z \gtrsim 6$ the number reaches almost double
the standard Press-Schechter prediction.

\citet{vandenbosch} has shown that there is a difference
between the average mass accretion history of haloes in simulations
and that derived from the semi-analytic merging history trees
\citep{somervillekolatt}. In particular the {\it N}-body simulations suggest
an earlier formation of haloes than is inferred from semi-analytic
trees. Earlier formation epochs are reminiscent of higher merging
activity towards high redshifts, which is the case in the moving
barrier model. In this context it is worth pointing out that our expressions
for conditional mass function could be used to generate ``improvised''
merging history trees.
The form of equation \ref{eq:final_cond} suggests that the task of drawing
progenitors using the given expression may not be as easy as in the
case of a constant barrier model. However, the analysis in section \ref{sec3.4}
indicates that when $\Delta a/\sqrt{t_1} \rightarrow 0 $ one can
use a considerably simplified expression for the conditional
distribution $f(t_2|t_1)$ (equation \ref{asympcond}). Therefore, for a
given present day halo with mass $M_1$ (corresponding to $t_1$) one
can choose an appropriately small $\Delta a$ and use equation
\ref{asympcond} to draw the progenitor masses. Alternatively we have
pointed out that the two opposite limits for the conditional mass
function are the constant barrier distribution and the distribution
obtained by simple re-scaling of the unconditional mass function. 
Hence the merging history trees in the moving barrier model would be
constrained by these limiting distributions. In a forthcoming work
this issue will be investigated in further detail.

\section*{acknowledgements}

The research of AM at Oxford was supported by the Eddie Dinshaw
foundation, Balliol College. RR was supported by NSF grant DMR-0207106.
We would like to thank Joe Silk, Julien Devriendt and James Taylor 
for their useful comments on the present work. 
We are grateful to Alessandro Marconi for explaining the
k-corrections for quasars. We also thank Adrian Jenkins and Ravi
Sheth for discussing various aspects of halo mass function in {\it N}-body
simulations, and Alexander Knebe and Darren Reed for making available
the outputs from their {\it N}-body simulations for comparison.

\appendix

\section{Parabolic cylinder functions}
\label{appendixa}

In this appendix, we discuss some standard
results pertaining to parabolic cylinder functions and their
integrals.
Parabolic cylinder functions are the solutions of the differential
equation \ref{pcdiffeq}. An integral representation of these
functions for $v \ge -1$ is given as \citep{erdelyi}
\be
U_v(x) = \sqrt{\frac{2}{\pi}} e^{x^2/4}  \int_0^{\infty}dt \ 
e^{- t^2/2}
 t^{v} \cos\left( x  t - \frac{v  \pi}{2} \right). 
\ee

For large $v$, the asymptotic form of these functions is
\bea
U_v(x) &=& \sqrt{2} \left(\frac{v}{e}\right)^{v/2} \left[\cos\left(x
  \sqrt{v} - \frac{v \pi}{2} \right) \right. \nonumber \\
& & \left. + \frac{x(x^2 - 6)}{24 \sqrt{v}} \sin \left(x \sqrt{v} - \frac{v
    \pi}{2}\right) +  O (1/v) \right]. 
\label{parabolic_asymp}
\eea
For the problem discussed in the paper, the boundary condition is
$U_v(-b/\sqrt{D})=0$. For large $v$, we can keep just the first term in 
equation \ref{parabolic_asymp} and estimate the eigenvalues to be given by
\be
\frac{\pi v}{2} + \frac{b \sqrt{v}}{\sqrt{D}} = (2 n + 1) \frac{\pi}{2},
\label{dn}
\ee
where $n$ is a large positive integer. 
Solving, we obtain
\be
v_n = 2n  - \frac{2 \sqrt{2}  b}{\pi \sqrt{D} }\sqrt{n} +
\frac{2  b^2}{\pi^2D} +(2 k-1) + O\left(\frac{1}{\sqrt{n}}\right).
\ee
Here, $k$ is an arbitrary integer.

To compute the asymptotic behavior of $U'_v(x)$ we note the recursion
relation $2  U'_v(x) = x U_v (x) - 2  U_{v + 1} (x)$. Thus for $x =
-b/\sqrt{D}$ we have $U'_v(-b/\sqrt{D}) = - U_{v+1}(x)$. Using
equation \ref{parabolic_asymp} we can then write the large $v$ asymptotic
form for $U'_v(-b/\sqrt{D})$ as 
\be
U'_v \left(\!\frac{-b}{\sqrt{D}}\right) \!=\!
\sqrt{2 v} \left( \frac{v}{e}\right)^{v/2}
\!\!\sin \!\left(\!  \frac{v \pi}{2}\! +\! \frac{b \sqrt{v}}{\sqrt{D}}\right)
\!\left[1 + O(1/v)\right].
\ee

In order to compute the large $v$ asymptotic form of the integral 
$I_v(-b/\sqrt{D})$ we first note that \citep{erdelyi}
\be
\int_0^\infty U_v^2(y)  dy = \frac{\sqrt{\pi}}{2^{3/2}} \left[\frac{\psi(1/2
  - v/2) - \psi(-v/2)}{\Gamma(-v)}\right],
\ee
where $\Gamma(x)$ is a gamma function and $\psi(x)$ is a logarithmic
derivative of the gamma function, defined as $\psi(x) =
\Gamma'(x)/\Gamma(x)$. Using a set of standard results [for e.g. given
in the appendix of \citet{benderorszag}] we  obtain the large $v$ behavior
of this integral as
\be
\int_0^\infty U_v^2(y)  dy = \pi \sqrt{e} \left(\frac{v}{e}\right)^{v
+ 1/2} [1 + O(1/v)].
\ee
Now in the remaining integral $\int_{-b/\sqrt{D}}^0 U_v^2(y)  dy$ we
use the large $v$ asymptotic form of the function $U_v(y)$ and
obtain the final integral as
\be
I_v(-b/\sqrt{D}) = \pi \sqrt{v} \left( \frac{v}{e}\right)^{v }
\left[1 + \frac{b}{\pi \sqrt{v  D}} + O(1/v) \right].
\ee


\begin{thebibliography}{999}

\bibitem[\protect\citeauthoryear{Bardeen et. al.}{1986}]{bbks}
Bardeen J. M., Bond J. R., Kaiser N., Szalay A. S., \ApJ{304}{15}{1986} (BBKS)

\bibitem[\protect\citeauthoryear{Bender \& Orszag}{1978}]{benderorszag}
Bender C. M., Orszag S. A., {\it Advanced Mathematical Methods for Scientists
and Engineers}, 1978, McGraw-Hill Book Company, New York


\bibitem[\protect\citeauthoryear{Bond et. al.}{1991}]{bondetal}
Bond J. R., Cole S., Efstathiou G., Kaiser N., 1991, ApJ, 379, 440 (BCEK)

\bibitem[\protect\citeauthoryear{Doroshkevich}{1970}]{doroshkevich}
Doroshkevich A. G., 1970, Astrofizika, 3, 175

\bibitem[\protect\citeauthoryear{Del Popolo \& Gambera}{1998}]{delpopoloetal} 
Del Popolo A., Gambera M., \AAA{337}{96}{1998}

\bibitem[\protect\citeauthoryear{Del Popolo}{2002}]{delpopolo} 
Del Popolo A., \MNRAS{337}{529}{2002}

\bibitem[\protect\citeauthoryear{Erdelyi}{1953}]{erdelyi}
Erdelyi A., ed., {\it Higher Transcendental Functions}, Volume 2,
1953, McGraw-Hill Book Company, New York 

\bibitem[\protect\citeauthoryear{Fan et al.}{2003}]{fanetal}
Fan X., et al., \AJ{125}{1649}{2003}

\bibitem[\protect\citeauthoryear{Feller}{1970}]{feller} 
Feller W., {\it An introduction to probability theory
and its applications}, 1970, Wiley, New York

\bibitem[\protect\citeauthoryear{Gelb \& 
Bertschinger}{1994}]{gelbbertschinger} 
Gelb J., Bertschinger E., \ApJ{436}{467}{1994}

\bibitem[\protect\citeauthoryear{Glazebrook et al.}{2004}]{glazebrooketal}
Glazebrook K., et al., \Nature{340}{181}{2004}

\bibitem[\protect\citeauthoryear{Grassberger}{2002}]{grassberger} 
Grassberger P., 2002, Comp. Phys. Commun., 147, 64 

\bibitem[\protect\citeauthoryear{Haiman \& Loeb}{1998}]{haimanloeb}
Haiman Z., Loeb A., \ApJ{503}{505}{1998}

\bibitem[Jenkins et al.(1998)]{jenkinsetal1} 
Jenkins A., Frenk C. S., Pearce F. R., Thomas P. A., Colberg J. M.,
White S. D. M., Couchman H. M. P., Peacock J. A., Efstathiou G.,
Nelson A. H., \ApJ{499}{20}{1998} 

\bibitem[Jenkins et al.(2001)]{jenkinsetal2} 
Jenkins A., Frenk C. S., White S. D. M., Colberg J. M., Cole S.,
Evrard A. E., Couchman H. M. P. \& Yoshida N., \MNRAS{321}{372}{2001} 

\bibitem[\protect\citeauthoryear{Kauffmann, White \& 
Guiderdoni}{1993}]{kauffmannwhiteguiderdoni} 
Kauffmann G., White S. D. M., Guiderdoni B., 1993, MNRAS, 264, 201

\bibitem[\protect\citeauthoryear{Krapivsky \& Redner}{1996}]{krapivskyredner} 
Krapivsky P. L., Redner S., 1996, Am. J. Phys., 64, 546.

\bibitem[\protect\citeauthoryear{Lacey \& Cole}{1994}]{laceycole2}
Lacey C., Cole, S., \MNRAS{271}{676}{1994}

\bibitem[\protect\citeauthoryear{Lacey \& Cole}{1993}]{laceycole1}
Lacey C., Cole, S.,  \MNRAS{262}{627}{1993} 

\bibitem[\protect\citeauthoryear{Mahmood, 
Devriendt \& Silk}{2004}]{mahmoodetal} 
Mahmood A., Devriendt J. E. G., Silk J., preprint (astro-ph/0401003)

\bibitem[\protect\citeauthoryear{Marconi et al.}{2004}]{marconietal04}
Marconi A., Risaliti G., Gilli R., Hunt L.K., Maiolino R., Salvati M.,
\MNRAS{169}{185}{2004}

\bibitem[Mo \& White(1996)]{mowhite} 
Mo H. J.~\& White S. D. M., \MNRAS{282}{347}{1996} 

\bibitem[\protect\citeauthoryear{Monaco}{1997a}]{monacoa} 
Monaco P., \ApJ{287}{753}{1997a} 

\bibitem[\protect\citeauthoryear{Monaco}{1997b}]{monacob} 
Monaco P., \ApJ{290}{439}{1997b}

\bibitem[Press \& Schechter(1974)]{pressschechter} 
Press W. H.~\& Schechter P., \ApJ{187}{425}{1974} 

\bibitem[\protect\citeauthoryear{Sheth \& Tormen}{1999}]{shethtormen1}
Sheth R. K., Tormen G.,  \MNRAS{308}{119}{1999}

\bibitem[\protect\citeauthoryear{Sheth \& Lemson}{1999}]{shethlemson}
Sheth R. K., Lemson G., \MNRAS{304}{767}{1999}

\bibitem[\protect\citeauthoryear{Sheth, Mo \& Tormen}{2001}]{shethmotormen}
Sheth R. K., Mo H. J., Tormen G., \MNRAS{323}{1}{2001}

\bibitem[\protect\citeauthoryear{Sheth \& Tormen}{2002}]{shethtormen2}
Sheth R. K., Tormen G.,  \MNRAS{329}{61}{2002}

\bibitem[\protect\citeauthoryear{Somerville \& Kolatt}{1999}]{somervillekolatt}
Somerville R., Kolatt T. S., \MNRAS{305}{1}{1999}

\bibitem[\protect\citeauthoryear{Tormen}{1998}]{tormen1998}
Tormen G., \MNRAS{297}{648}{1998}

\bibitem[\protect\citeauthoryear{van den Bosch}{2002}]{vandenbosch}
van den Bosch F., \MNRAS{331}{98}{2002}

\bibitem[\protect\citeauthoryear{Wyithe \& Loeb}{2002}]{wyitheloeb02}
Wyithe J. S. B., Loeb A., \ApJ{581}{886}{2002}

\bibitem[\protect\citeauthoryear{Wyithe \& Loeb}{2003}]{wyitheloeb}
Wyithe J. S. B., Loeb A., \ApJ{595}{614}{2003}

\end{thebibliography}
\end{document}